\documentstyle[preprint,aps,epsf]{revtex}
\include{epsf}
 \def\odiff#1#2#3{\frac{d^{#3}#1}{d#2^{#3}}}
\tighten
\begin{document}
\draft
\title{Ultracold collisions of metastable helium atoms}
\author{P.J. Leo}
\address{Atomic Physics Division, National Institute of Standards
and Technology, Gaithersburg, Maryland 20899}
\author{V. Venturi\thanks{Present address: Department of Computing Science, 
Glasgow University, Glasgow G12 8QQ, UK} and I.B. Whittingham }
\address{School of Mathematical and Physical Sciences,
James Cook University, Townsville 4811, Australia}
\author{J. F. Babb} 
\address{Institute for Theoretical Atomic and Molecular Physics,\\
Harvard-Smithsonian Center for Astrophysics,\\ 
60 Garden Street, Cambridge, MA 02138 }
\date{\today}
\maketitle

\begin{abstract}
We report scattering lengths for the ${}^1 \Sigma_g^{+}$, 
${}^3 \Sigma_u^{+}$ and ${}^5 \Sigma_g^{+}$ adiabatic molecular 
potentials relevant to collisions of 
two metastable $2\;{}^3S$ 
helium atoms as a function
of the uncertainty in these potentials. 
These scattering lengths are used to calculate experimentally observable 
scattering lengths, elastic cross sections and inelastic rates 
for any combination of states of the colliding atoms, 
at temperatures where the Wigner threshold approximation is
valid. 
\end{abstract}


\narrowtext

\section{Introduction} \label{intro}

Metastable helium has been the subject of many experimental investigations at 
cold and ultracold temperatures. 
These include various methods of laser cooling \cite{Aspectetal:1988,%
Aspectetal:1989,Morita&Kumakura:1991,Lawalletal:1994,Lawalletal:1995,%
Rooijakkersetal:1995,Rooijakkersetal:1997b,Schumannetal:1999} and 
trapping \cite{Metcalf:1989,Westbrooketal:1991,Kumakura&Morita:1992,%
Vassen:1997,Rooijakkersetal:1997a}, production of an intense beam 
\cite{Aspectetal:1990,Rooijakkersetal:1996,Hoogerlandetal:1996b}, optical
collisions in magneto-optical traps and measurements of two-body trap
loss rates, including that due to Penning ionization \cite{Bardouetal:1992,%
Mastwijketal:1998a,Mastwijketal:1998b,Woestenenketal:1999,%
Kumakura&Morita:1999,Toletal:1999,Browaeysetal:2000,%
Herschbachetal:2000a,Pereiraetal:2001a}, photoassociation spectroscopy 
\cite{Herschbachetal:2000b} and magnetostatic trapping \cite{Nowaketal:2000}.
There have also been several theoretical
studies~\cite{Julienne&Mies:1989,Williams&Julienne:1992,Shlyapnikovetal:1994,%
Fedichevetal:1996a,Venturietal:1999,Venturi&Whittingham:2000}. Much of this
interest has been stimulated by the prospect of obtaining a Bose-Einstein
condensate 
with spin-polarized metastable helium $2\;{}^3S$
atoms~\cite{Metcalf:1989,Vassen:1997,%
Toletal:1999,Nowaketal:2000,Pereiraetal:2001b,%
Fedichevetal:1996a,Venturietal:1999}; 
a quest successfully realized very recently \cite{Robertetal:2001,%
Pereiraetal:2001c}.

The Penning ionization (PI) and associative
ionization (AI) processes,
\begin{equation}
\mathrm{He}^{\ast} + \mathrm{He}^{\ast} \rightarrow 
\left\{ \begin{array}{ll} \mathrm{He} + \mathrm{He}^{+} + e^{-} 
\quad & (\mathrm{PI}) \\
\mathrm{He}^{+}_2 + e^{-} \quad & (\mathrm{AI}) \,  \end{array} \right.
\end{equation}
have high threshold rates in an unpolarized gas and limit the achievable
density of trapped atoms.
Here we denote a $\mathrm{He}(2\;{}^3S$) atom by the
symbol $\mathrm{He}^{\ast}$.
However, these autoionization processes are 
spin-forbidden and 
suppressed~\cite{Venturietal:1999,Venturi&Whittingham:2000} from the 
spin-polarized state and only via 
the weak spin-dipole interaction can such processes occur.
Consequently,  a sufficient number of spin-polarized metastable
atoms should remain trapped. In addition, 
the scattering length associated with the
quintet potential, which controls the collision dynamics of spin-polarized
metastable helium atoms, is predicted to be large and positive, a
necessary requirement for a stable Bose-Einstein condensate.
Although some theoretical studies~\cite{Shlyapnikovetal:1994,%
Fedichevetal:1996a,Venturietal:1999} have estimated the scattering length
associated with the quintet potential to be large and positive, no detailed
study of the scattering lengths for metastable helium has been previously
undertaken. 

In this present investigation we calculate not only the possible
ranges of values for the scattering lengths directly associated with the
molecular potentials, but also experimentally observable scattering
lengths, elastic cross sections and inelastic rate constants 
over a range of scattering lengths and temperatures 
for collisions of
metastable helium atoms in the presence of a magnetic field.
Measurement of cross sections or rates should then provide information on
the scattering lengths and hence the potentials.

For this theoretical investigation we have chosen to simulate the Penning and
associative ionization processes that occur at small internuclear separations
from the singlet and triplet molecular states by a complex optical potential.
The complex interaction potentials then have the form
${}^{2S+1} V(R) - \case{1}{2}i \, {}^{2S+1}\Gamma(R)$,
where $R$ is the internuclear separation of the two atoms,
${}^{2S+1}V(R)$ is the usual adiabatic molecular potential for the
molecular state ${}^{2S+1} \Sigma_{g,u}^{+}$ with total spin $S$, and
${}^{2S+1}\Gamma(R)$ is the corresponding total autoionization width
representing flux loss due to the ionization processes. Since the Penning and
associative ionization processes are spin-forbidden from the quintet state,
${}^{5}\Gamma(R)=0$.

In the absence of spectroscopic data which could be used to obtain high
accuracy potentials, 
the adiabatic molecular potentials required in this investigation for the
${}^1 \Sigma_g^{+}$, ${}^3 \Sigma_u^{+}$ and ${}^5 \Sigma_g^{+}$ molecular
states were constructed using data from various sources. The long-range
interaction potential was described by a multipole expansion of the form
$-C_6/R^6-C_8/R^8-C_{10}/R^{10}$ using the most accurate dispersion
coefficients available for the He$(2\,^3\!S)$--He$(2\,^3\!S)$ interaction
($C_{6}=3276.680,\;C_{8}=210566.55,\;C_{10}=21786760$ au)
\cite{Yan&Babb:1998}. The short-range ${}^1 \Sigma_g^{+}$ and
${}^3 \Sigma_u^{+}$ molecular potentials and their corresponding
autoionization widths for Penning and associative ionization were obtained
from M\"{u}ller~\textit{et al.}~\cite{Mulleretal:1991}, while the short-range
${}^5 \Sigma_g^{+}$ molecular potential was taken from St\"{a}rck and
Meyer~\cite{Starck&Meyer:1994}. The ${}^5 \Sigma_g^{+}$ potential was reported
with an uncertainty of 0.5\% in the repulsive part of the potential and 1\%
in the attractive part of the potential.

The molecular potentials for metastable helium were constructed by fitting
the three short-range potentials smoothly onto the long-range dispersion
interaction around $\approx 20 \,a_{0}$ by interpolation through this
region using an Akima spline fitted to $R^{6}\;{}^{2S+1}V(R)$. 
The uncertainties in the short-range potentials, the procedure
used to connect these to the long-range potential, and the form used for the
autoionization widths lead to uncertainties in the scattering lengths
for the ${}^1 \Sigma_g^{+}$, ${}^3 \Sigma_u^{+}$ and ${}^5 \Sigma_g^{+}$
potentials and subsequently to the ultracold scattering properties of
metastable helium atoms.
To determine the extent of these uncertainties we vary the short-range
potentials by $\pm 2.5$\% for five different interaction potentials which use
different methods to connect the long-range and short-range potentials or
have different forms of the autoionization widths. 
Since there is no available experimental data that can be used to determine
the level of accuracy of these short-range potentials,
we have chosen to vary them
by more than their stated uncertainty to ensure that we
obtain conservative estimates for the range of scattering lengths.

The first of these potentials, labeled (A), uses the analytic short-range
${}^5 \Sigma_g^{+}$ potential fitted smoothly onto the long-range potential
at $R \approx 20\,a_0$. The numerical 
${}^1 \Sigma_g^{+}$ and ${}^3 \Sigma_u^{+}$
molecular potentials of Ref. \cite{Mulleretal:1991} are used for
$R<11.5\,a_0$ and for larger $R$, where the electronic structure calculations
become inaccurate, we replace the potentials by
${}^5 V (R)- V_{\mathrm{exch}} (R)$. 
The exchange term has the form \cite{Tangetal:1998}
$V_{\mathrm{exch}} (R) = A_{2S+1} R^\gamma \exp(-\beta R)$, 
where \cite{Tangetal:1998}
$\gamma=4.91249$, $\beta=1.183933$, $A_1=6.3245\times10^{-3}$ and
$A_3=4.6317\times10^{-3}$. The autoionization widths
${}^{2S+1} \Gamma_{\mathrm{M}} (R)$ of Ref. \cite{Mulleretal:1991} were used
to represent the Penning and associative ionization processes.

Potential (B) is identical to (A) except that the short-range
${}^5 \Sigma_g^{+}$ form is fitted to the long-range potential at
$R \approx 35\,a_0$. 
Potential (C) is identical to (A) except that the
exchange term  has the form 
$V_{\mathrm{exch}} (R) = A_{2S+1} \exp(-\beta R)$
with $\beta=0.704921$, $A_1=4.29808$ and $A_3=3.14764$. Potentials (D) and (E)
are identical to (A) but employ different forms for the autoionization widths.
The autoionization width $\Gamma_{\mathrm{GMS}} (R)=0.3 \exp(-R/1.086)$, given
by Garrison~{\it et al.\/}~\cite{Garrisonetal:1973}, is used in (D). This
autoionization width has a steeper exponential form which doesn't dampen at
small internuclear separations like ${}^{1} \Gamma_{\mathrm{M}} (R)$ or
${}^{3} \Gamma_{\mathrm{M}} (R)$. Potential (E) uses another alternative
form of the autoionization widths which was arbitrarily constructed
to assess the sensitivity of the calculated results to the form of
$\Gamma (R)$ and is given by:
\begin{equation}
\Gamma (R) = \left\{ 
\begin{array}{ll}
\Gamma_{\mathrm{GMS}} (R) + (R-6.5)^2 e^{-0.75 \, R}, & \mbox{for}
\, R \leq 6.5 \\
\Gamma_{\mathrm{GMS}} (R), & \mbox{for} \, R> 6.5 .
\end{array} \right.
\label{eq:uppi}
\end{equation}
All the molecular potentials considered have the same long-range form since
the uncertainties in the long-range multipole potential were found to have a 
negligible effect on the scattering lengths. 
The real parts of the potentials (A) to (E) with
unmodified short-range forms possess the same number of bound states,
calculated to be 28 for ${}^1 \Sigma_g^{+}$, 27 for  ${}^3 \Sigma_u^{+}$ 
and 15 for ${}^5 \Sigma_g^{+}$.

\section{Scattering lengths associated with the molecular potentials}

The scattering lengths for the ${}^1 \Sigma_g^{+}$, ${}^3\Sigma_u^{+}$ and
${}^5 \Sigma_g^{+}$ molecular potentials were obtained by solving a single
channel radial Schr\"{o}dinger equation of the form,
\begin{eqnarray}
&& \biggl\{ \odiff{}{R}{2} - \frac{l(l+1)}{R^2} - \left[ {}^{2S+1}V (R)
\right. \nonumber \\ && \qquad \qquad \qquad \left.
- \case{1}{2}i \, {}^{2S+1}\Gamma(R) \right] + k^2 \biggr\} u_{S,l} (k,R) = 0
\label{eq:sf1}
\end{eqnarray}
in the limit where $k \rightarrow 0$.
Here $k=\sqrt{2 \mu E / \hbar^2}$, $\mu$ is the reduced mass of the atomic
system, $E$ is the total energy of the system and
$l$ is the relative rotational angular momentum. As a result of the complex
interaction potential, the scattering equation (\ref{eq:sf1}) and the
wavefunctions $u_{S,l}(k,R)$ are complex. Solution of this equation allowing
for the non-unitarity of the Hamiltonian, and subsequent fitting to free-field
boundary conditions, provides a complex K-matrix and corresponding non-unitary
S-matrix ($S_S$), as described previously \cite{Venturietal:1999}. 
The complex phase shift $\eta_{S}$, 
defined by $S_{S}=\exp (i 2\eta_{S})$, can
then be used to calculate the complex scattering lengths  
$a_{2S+1}=a^{\mathrm{re}}_{2S+1} + i a^{\mathrm{im}}_{2S+1}$ for each
molecular state ${}^{2S+1} \Sigma_{g,u}^{+}$:
\begin{eqnarray}
a^{\mathrm{re}}_{2S+1}&=&-\frac{1}{2k} \tan^{-1}
\left(\frac{S_S^{\mathrm{im}}}{S_S^{\mathrm{re}}}\right)
\nonumber \\
a^{\mathrm{im}}_{2S+1}&=& - \frac{\ln \left(S_S S_S^\dagger\right)}{4k}
\label{leneqn} ,
\end{eqnarray}
where the scattering lengths are defined by $\eta_S=-k\; a_{2S+1}^{*}$ 
and the
superscripts `$\mathrm{re}$' and `$\mathrm{im}$' denote real and imaginary
components, respectively. This definition means that 
$+ia_{2S+1}^{\mathrm{im}}$ represents a loss process.

\begin{figure}[tb]
\leftline{\epsfxsize=5.5in\epsfbox{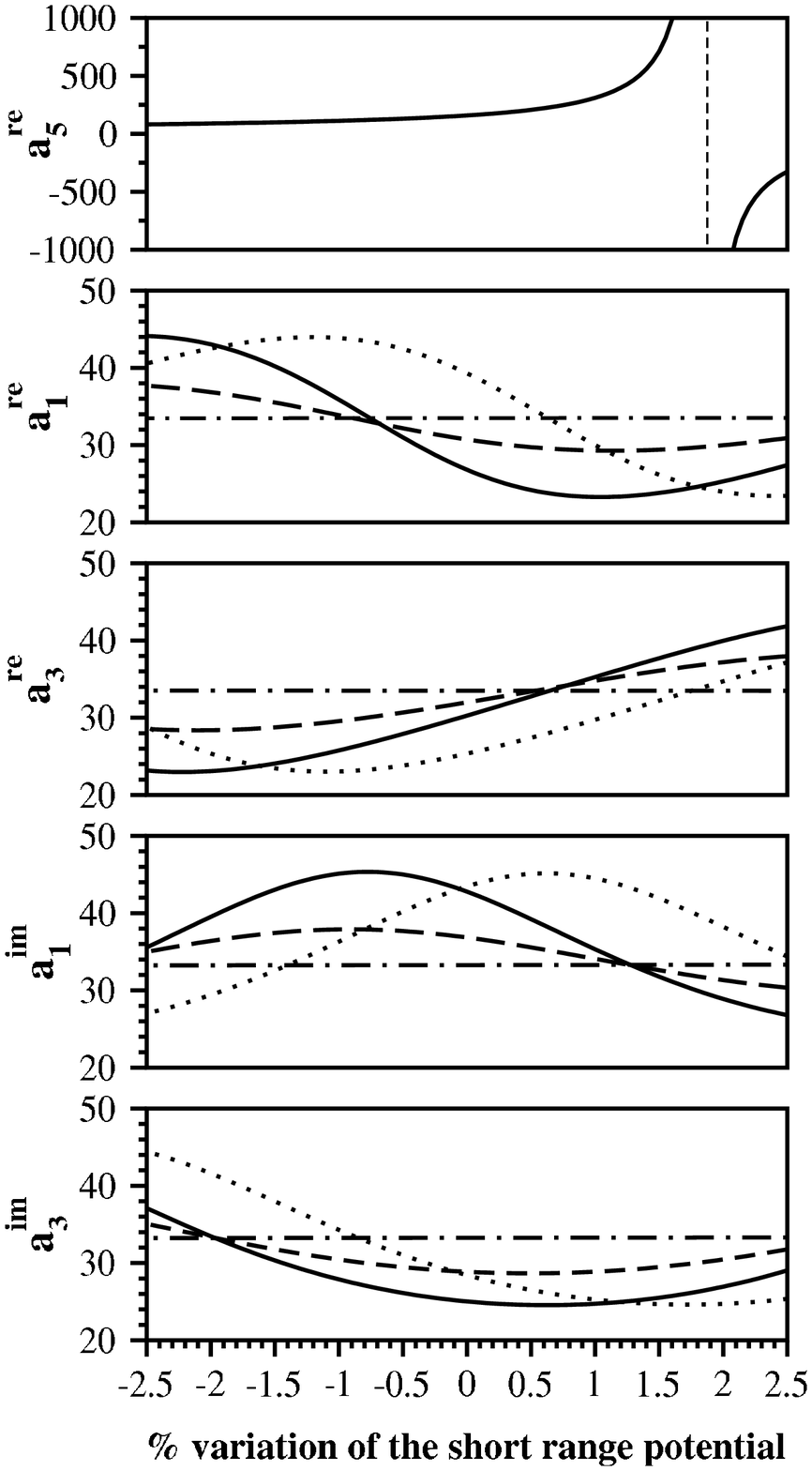}}
\caption{Real and imaginary  components of the scattering lengths  
$a_1$, $a_3$ and $a_5$ 
plotted against variation in the short-range potential. For $a_5$ the
five potential cases produced similar results and are 
encompassed by the solid curve with
a dashed line to denote the position of the resonance. For $a_1$ and $a_3$,
potentials (A) and (B) produced identical results denoted by (---), 
potential (C) by
($\cdot\cdot\cdot$), potential (D) by (- - -) and potential (E) 
by ($-\cdot-$).}
\label{a5a1a3}
\end{figure}

The scattering lengths for the three molecular states were calculated as
a function of the percentage variation in the corresponding short-range
molecular potential for the five potential cases (A) to (E) 
and are displayed in
Fig. \ref{a5a1a3}. 
For the $a_5$ scattering length only
the results for
potential (A) are plotted 
because the ${}^5\Sigma_g^{+}$ potentials are identical for
potential cases (A), (C), (D) and (E)  
and the results obtained with potential (B)
differed by less than 5\%. 
The $a_5$ scattering length has no imaginary component
since the Penning process is spin-forbidden from the $S=2$ molecular state.
Of particular interest is the resonance in
$a_5$ at a percentage variation of $\approx 1.875$ where the short-range
potential is made sufficiently shallow that a bound state is removed from
the ${}^5 \Sigma_g^{+}$ potential. 
For percentage variations $>1.875$ it is found that $a_5$ is negative
in contradiction to 
recent experimental
evidence that $a_5$ is large and
positive~\cite{Robertetal:2001,Pereiraetal:2001c}. 
With potentials (A) and (B)  the scattering lengths
$a_1$ and $a_3$ were nearly identical and are denoted by a single solid curve.

The scattering lengths associated with the molecular potentials are not
observable experimentally, with the exception of $a_5$, which is approximately
equal to the scattering length for the spin-polarized state. However, these
scattering lengths provide unique parameterization of the ${}^1 \Sigma_g^{+}$,
${}^3 \Sigma_u^{+}$ and ${}^5 \Sigma_g^{+}$ potentials, from which the
threshold scattering properties of metastable helium atoms can be obtained.
Of more practical interest are the scattering lengths for collisions between
atoms in given atomic states in the presence of a magnetic field.

\section{Collisions in the presence of a magnetic field}

To study collisions in the presence of a magnetic field a full multichannel
scattering calculation must be undertaken 
in which the total Hamiltonian describing the two-body collision
includes the spin-orbit, Zeeman and spin-dipole interactions in addition
to the usual radial and rotational kinetic energy operators of the two 
atoms and the electronic Hamiltonian of the quasimolecule formed
during the collision.
The details of such a
quantum-mechanical multichannel scattering model for metastable helium is
described elsewhere~\cite{Venturietal:1999}. In brief, we perform the present
calculations for atoms in initial atomic states $\alpha$ and $\beta$,
including both $s$ and $d$-waves, and calculate the full non-unitary S-matrix
which has elements $S_{\alpha,\beta,l;\alpha^\prime,\beta^\prime,l^\prime}$.
Here we let $\alpha$ and $\beta$ denote the atomic states 
$(s,m_s)$, where $m_s$ is the space-fixed projection of the spin $s$ for an
individual atom.

For collision energies up to $100\, \mu $K the contributions of entrance 
$p$ and
$d$-waves are negligible (note that due to symmetrization $p$-waves only 
contribute in collisions between atoms in different atomic states), so that
only the $s$-wave entrance channel $[\alpha,\beta],l=0$ needs to be
considered.
The  elastic cross section $\sigma^{\mathrm{el}}_{\alpha,\beta}$ and
inelastic rate $K^{\mathrm{inel}}_{\alpha,\beta}$ are then given
by~\cite{landau}
\begin{eqnarray}
\sigma^{\mathrm{el}}_{\alpha,\beta} &=& \frac{\pi}{k^2}
| 1-S_{\alpha,\beta,l=0;\alpha,\beta,l=0} |^2 \nonumber \\
K^{\mathrm{inel}}_{\alpha,\beta} &=&  \frac{v \pi}{k^2}
\left( 1-|S_{\alpha,\beta,l=0;\alpha,\beta,l=0} |^2 \right) \label{rateeqn} ,
\end{eqnarray}
where $v$ is the relative atomic velocity. In the Wigner threshold region
($ka<<1$) one can define the scattering lengths using
$\eta_{\alpha,\beta}=-k a_{\alpha,\beta}^{*}$ and obtain expressions for the
observable scattering lengths $a_{\alpha,\beta}$ by replacing $S_S$ with
the matrix element $S_{\alpha,\beta,l=0;\alpha,\beta,l=0 }$ in
Eq. (\ref{leneqn}). The elastic cross sections and inelastic rates can then
be obtained using
\begin{eqnarray}
\sigma^{\mathrm{el}}_{\alpha,\beta} &=& 4 \pi \left[
(a^{\mathrm{re}}_{\alpha,\beta})^2 +(a^{\mathrm{im}}_{\alpha,\beta})^2
\right] \nonumber \\
K^{\mathrm{inel}}_{\alpha,\beta} &=&  4 \pi\, a^{\mathrm{im}}_{\alpha,\beta}
/k \label{rateeqnL}.
\end{eqnarray}
The inelastic rate $K^{\mathrm{inel}}_{\alpha,\beta}$ includes both
contributions from the flux loss due to Penning ionization and that due to
the atoms exiting in different atomic states. Since we calculate the full
S-matrix, the contributions of these two processes can be easily separated.  
We note that for $(1,1)+(1,1),\; (1,1)+(1,0),\; (1,-1)+(1,-1)$ or
$(1,-1)+(1,0)$ collisions, where the total projection of the spin ($M$) is 
non-zero,
the collision is dominated by the ${}^5 \Sigma_g^{+}$ potential. 
This is because parity considerations associate the 
${}^3 \Sigma_u^{+}$ potential 
with odd partial waves, and cold collisions are dominated by $s$ wave
collisions, and the ${}^1 \Sigma_g^{+}$ potential can only contribute 
when $M=0$. Hence
inelastic 
processes can only occur via the weak relativistic spin-dipole interaction. 
The scattering lengths for these states are then almost identical to $a_5$
but with a small imaginary component.
The properties of $(1,1)+(1,1)$ collisions were investigated in detail in 
a previous paper~\cite{Venturietal:1999}.
The inelastic rates for $(1,0)+(1,0)$ and $(1,1)+(1,-1)$ collisions, from
which ionization can occur directly via strong exchange forces, are much
larger and dominate the total inelastic rate for an unpolarized gas. 

The $(1,0)+(1,0)$ and $(1,1)+(1,-1)$ inelastic rates contain two different
contributions. The first is due to exothermic inelastic processes which
includes the Penning rate $K^{\text{P}}_{\alpha,\beta}$ 
and the much smaller collision rate for
exothermic fine-structure changing collisions $K^{\text{ex}}_{\alpha,\beta}$. 
The second is the rate for degenerate fine-structure changing collisions
$K^{\text{deg}}_{\alpha,\beta}$.
For example, in ultracold $(1,0)+(1,0)$ collisions, the entrance channel
$[(1,0)+(1,0)], l=0$ can decay exothermically to the three channels
$[(1,-1)+(1,-1)],l=2$; $[(1,0)+(1,-1)],l=2$ and the Penning channel, and to
the two degenerate channels $[(1,1)+(1,-1)],l=0$ and $[(1,1)+(1,-1)],l=2$. 
The flux loss to the degenerate $d$-wave exit channels or exothermic 
$d$-wave exit channels (ie $K^{\text{P}}$) only occurs via weak spin-dipole
forces  and is  at least 3 orders of magnitude
smaller than that lost to the Penning channel or to degenerate 
$l=0$ exit channels that occurs 
through strong exchange forces. 
Importantly, exothermic and degenerate inelastic processes exhibit different
threshold properties. Exothermic inelastic rates tend to a constant in the
Wigner threshold region whereas degenerate inelastic rates fall off as
$1/k$ since, as for elastic 
processes, 
the incident and final wave number are
identical. To represent these separate threshold behaviors in the inelastic
rates, we write
$a^{\mathrm{im}}_{\alpha,\beta}=a^{\mathrm{im\,ex}}_{\alpha,\beta} + k \,
a^{\mathrm{im\,deg}}_{\alpha,\beta}$.
The slope and intercept of
$\ln (S_{\alpha,\beta,l=0;\alpha,\beta,l=0}
S^\dagger_{\alpha,\beta,l=0;\alpha,\beta,l=0})/4k$
vs $k$, for $k$ in the Wigner threshold region, then gives the degenerate and
exothermic scattering lengths $a^{\mathrm{im\,deg}}_{\alpha,\beta}$ and
$a^{\mathrm{im\,ex}}_{\alpha,\beta}$, respectively.

We have calculated these imaginary and the real scattering lengths for all
possible collision processes in spin-polarized metastable helium for the five
different potentials under investigation. From these calculated scattering
lengths one can use Eq. (\ref{rateeqnL}) to calculate the partial rates 
or the total rates in an unpolarized gas at temperatures where the
Wigner threshold approximation is valid.
The scattering lengths are calculated assuming a magnetic field of 10 Gauss,
however we find only a weak dependence on the magnetic field and results for
fields in the range 0 to 20 Gauss differ by less than 1\%.
The scattering lengths can be used to calculate the rates and cross sections
up to typically $\approx 100\,\mu$K, except where the scattering lengths
become $>1000\,a_0$.

Scattering lengths for $(1,0)+(1,0)$ and $(1,1)+(1,-1)$ collisions (with 
$s$
and $d$-waves) depend on both the ${}^5 \Sigma_g^{+}$ and ${}^1 \Sigma_g^{+}$
potentials and so their scattering lengths are a function of both the 
percentage
variation of the short-range ${}^5 \Sigma_g^{+}$ and  ${}^1 \Sigma_g^{+}$ 
potentials for potential cases (A) to (E). However, we find that for a fixed
percentage variation  of  the ${}^5 \Sigma_g^{+}$ potential, the uncertainty in the
scattering lengths induced by varying the short-range ${}^1 \Sigma_g^{+}$
potential by $\pm 2.5\%$ is similar to that calculated by fixing the percentage
variation in the ${}^1 \Sigma_g^{+}$ potential to zero and using the five
different potential cases (A) to (E).
In all instances the percentage variation in the ${}^5 \Sigma_g^{+}$ potential
has the largest effect on the scattering lengths and resulting rates.
The $(1,1)+(1,1),\;(1,1)+(1,0),\;(1,-1)+(1,-1)$ or $(1,-1)+(1,0)$ 
interactions depend only weakly on the singlet potential via the weak
relativistic spin-dipole interaction  and we find that varying the 
short-range ${}^1 \Sigma_g^{+}$ potential for these collisions produces
negligible changes in the scattering length.
Therefore we only report scattering lengths as a function of the
percentage variation in the ${}^5 \Sigma_g^{+}$ potential for potential
cases (A) to (E), with the understanding that similar uncertainties result in
the $(1,0)+(1,0)$ and $(1,1)+(1,-1)$ scattering lengths by varying the
short-range singlet potential.

\begin{figure}[tb]
\centerline{\epsfxsize=5.0in\epsfbox{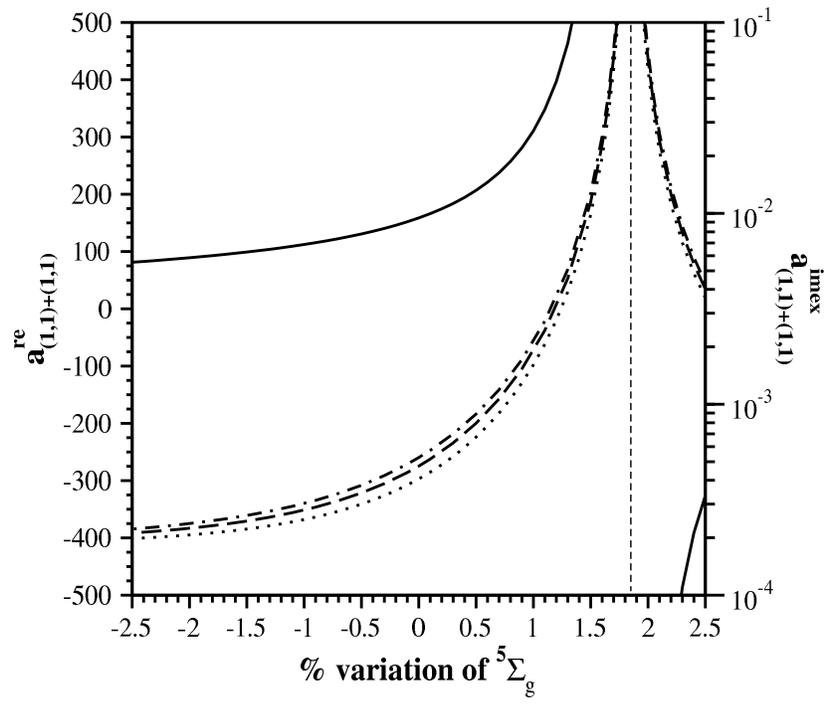}}
\caption{Complex scattering lengths
for $(1,1)+(1,1)$ collisions. The solid line includes the real components
of the scattering lengths
obtained from all five potential cases. The dashed and dotted lines 
give the imaginary components of the
scattering lengths. Results for potentials (A), (B) and (C) are given by 
($\cdot\cdot\cdot$), potential (D) by (- - -) and potential (E) 
by ($-\cdot-$).}
\label{a5eim}
\end{figure}

\begin{figure}[tb]
\centerline{\epsfxsize=5.0in\epsfbox{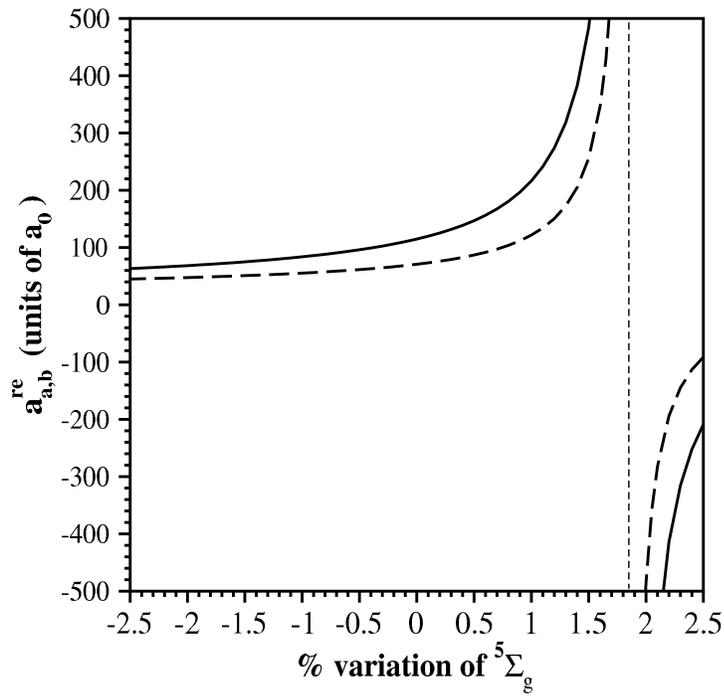}}
\caption{Real components of the scattering lengths
for $(1,0)+(1,0)$  and $(1,-1)+(1,1)$ collisions.
The solid line represents the results for $(1,0)+(1,0)$ collisions 
for all five  potential cases, the dashed line includes results
for $(1,-1)+(1,1)$ collisions for all five potential cases. }
\label{ars}
\end{figure}

\begin{figure}[tb]
\centerline{\epsfxsize=5.0in\epsfbox{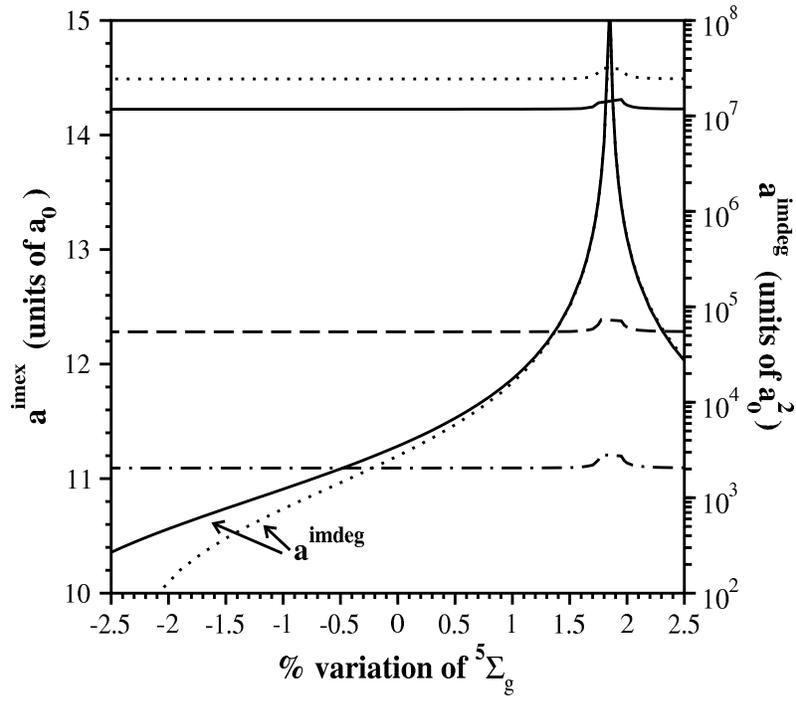}}
\caption{Imaginary components of the scattering lengths
for $(1,0)+(1,0)$ and $(1,-1)+(1,1)$ collisions. 
The near horizontal lines are for $(1,0)+(1,0)$ collisions with results for
potentials (A) and (B) encompassed by the solid curve, 
potential (C) by
($\cdot\cdot\cdot$), potential (D) by (- - -) and potential (E) 
by ($-\cdot-$). Note that 
$2a^{\mathrm{im\,ex}}_{(1,0),(1,0)}=a^{\mathrm{im\,ex}}_{(1,-1),(1,1)}$. 
The values of $a^{\mathrm{im\,deg}}$ for $(1,0)+(1,0)$ and 
$(1,-1)+(1,1)$ collisions are equal and
results for potentials (A), (B), (D) and (E) are encompassed 
by the solid curve and
those for potential (C) by the dotted curve.}
\label{ais}
\end{figure}

Figures \ref{a5eim}, \ref{ars} and \ref{ais} show the real and imaginary 
components of the
scattering lengths for $(1,1)+(1,1),\; (1,0)+(1,0)$ and $(1,1)+(1,-1)$
collisions. The real scattering lengths all possess a resonance in the region 
where a bound state is removed from the ${}^5 \Sigma_g^{+}$ potential and
$a_5$ goes through $\pm \infty$. Similar plots exist for
$(1,1)+(1,0),\; (1,-1)+(1,-1)$ and $ (1,-1)+(1,0)$ collisions but are almost
identical to that shown in Fig.~\ref{a5eim} 
for $(1,1)+(1,1)$ since all are dominated by the
${}^5 \Sigma_g^{+}$ potential. 
The underlying ${}^5 \Sigma_g^{+}$ potentials are identical for potential cases
(A), (C), (D) and (E) and we found that $a^{\mathrm{re}}_{(1,1),(1,1)}$ calculated 
with these
potentials   differed from those obtained using potential  (B) by less than
2\%. These small differences are not observable on the scale used in Fig.
\ref{a5eim} and so for clarity a single solid curve is used to represent
$a^{\mathrm{re}}_{(1,1),(1,1)}$ for the five potential cases. Similarly for 
$a^{\mathrm{imex}}_{(1,1),(1,1)}$ the results were identical except for cases (D) and (E)
where different forms of the 
autoionization widths were used, and so we show only
results for (A), (D) and (E) potential cases. 
We note that imaginary scattering lengths for
collisions where the total spin projection is non-zero
possess no degenerate component and the exothermic contributions are
negligible when compared to those for $(1,0)+(1,0)$ and 
$(1,1)+(1,-1)$ collisions
where Penning ionization can occur via exchange forces. 

For $(1,0)+(1,0)$ and $(1,1)+(1,-1)$ collisions
the values of $a^{\mathrm{re}}$ in  Fig.~\ref{ars} were  almost
identical for the five potential cases and are 
represented by a single solid curve
for $(1,0)+(1,0)$ and a dashed curve for $(1,1)+(1,-1)$. 
In Fig.~\ref{ais} we show 
$a^{\mathrm{im\,ex}}$ and $a^{\mathrm{im\,deg}}$ for these collisions. 
The scattering lengths $a^{\mathrm{im\,ex}}$, 
which measure 
$K^{\text{P}}_{\alpha,\beta}+K^{\text{ex}}$, 
are independent of the percentage variation in the
${}^5 \Sigma_g^{+}$ potential and thus $a_5$, except very near the $a_5$
resonance where the contribution from $K^{\text{ex}}_{\alpha,\beta}$ 
is no longer
negligible and a small increase in $a^{\mathrm{im\,ex}}_{\alpha,\beta}$ is
observable. Therefore, the measurement of the ionization signal from trapped
metastable helium atoms does not provide information about $a_5$, the
parameter which is required to make predictions of the formation or properties
of a Bose condensate of spin-polarized metastable helium atoms. 
If $K^{\text{ex}}_{\alpha,\beta}$ is neglected then a 
simple examination of the
Hamiltonian shows that 
$2\,K^{\text{P}}_{(1,0),(1,0)}=K^{\text{P}}_{(1,1),(1,-1)}$. We have verified
that this relation is valid to better than 1\% and so in Fig.~\ref{ais} we plot
results for $a^{\mathrm{im\,ex}}_{(1,0),(1,0)}$ for the five potential cases 
with the understanding that 
$2\,a^{\mathrm{im\,ex}}_{(1,0),(1,0)}=a^{\mathrm{im\,ex}}_{(1,1),(1,-1)}$.

The  curves labeled $a^{\mathrm{im\,deg}}$ 
in  Fig.~\ref{ais} provide the
degenerate temperature-dependent inelastic
rates for either $(1,0)+(1,0) \rightarrow (1,1)+(1,-1)$ or $(1,1)+(1,-1)
\rightarrow (1,0)+(1,0)$.  These equal, 
exchange-dominated rates strongly mix the
$(1,1)$, $(1,0)$ and $(1,-1)$ atoms and are equal to, or larger than, $K^{\text{P}}$ at
temperatures greater than $500\, \mu $K or when the quintet
potential is near resonance. Of the potentials tested only those with very
different exchange terms provided significantly different results and  consequently  
$a^{\mathrm{im\,deg}}$  for potentials (A), (B), (D) 
and (E) were nearly identical.
For convenience only $a^{\mathrm{im\,deg}}_{(1,0),(1,0)}$ 
for potentials (A) and (C)
have been plotted in Fig.\ref{ais}.

\begin{figure}[tb]
\centerline{\epsfxsize=5.0in\epsfbox{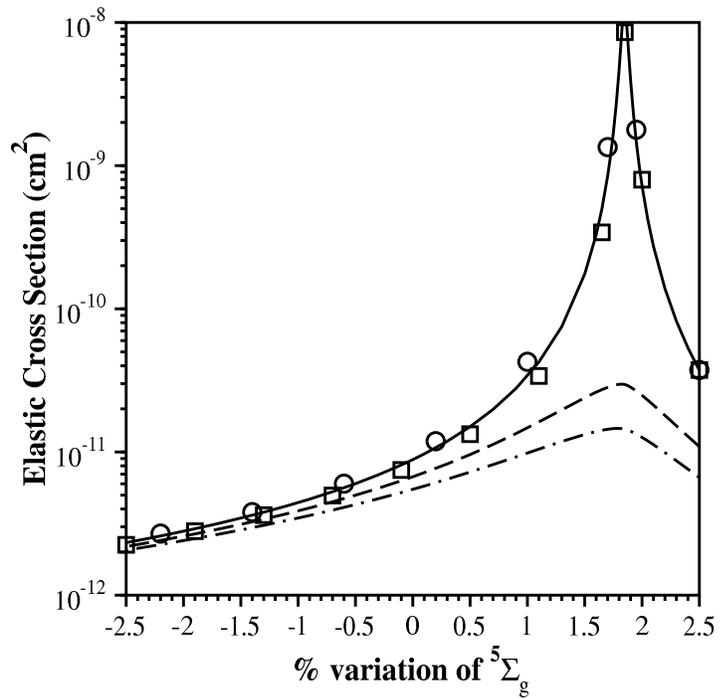}}
\caption{Thermally-averaged elastic cross section for $(1,1)+(1,1)$ 
collisions with potential (A) at various temperatures denoted by (---) for 
$1\,\mu$K, (- - -) for $500\,\mu$K, ($-\cdot-$) for  $1000\,\mu$K. 
Results for $1\,\mu$K and $500\,\mu$K calculated using the scattering lengths are denoted by 
$\Box $ and $\bigcirc $  respectively.}
\label{totalcce}
\end{figure}

\begin{figure}[tb]
\centerline{\epsfxsize=5.0in\epsfbox{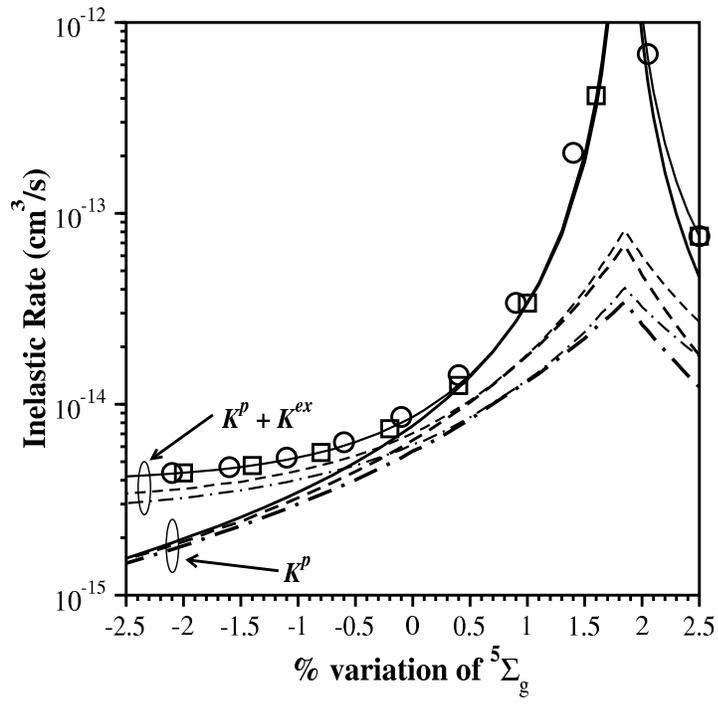}}
\caption{Thermally-averaged inelastic rates for $(1,1)+(1,1)$ 
collisions for potential (A) with curves  and symbols labeled 
using the same scheme 
as in   Fig. \ref{totalcce}. 
Thick lines labeled $K^{\mathrm{P}}$ denote the Penning 
rate $K^{\mathrm{P}}$ and the thinner lines labeled 
$K^{\mathrm{P}}+K^{\mathrm{ex}}$ give the
total inelastic rate.}
\label{totalccin}
\end{figure}

\begin{figure}[tb]
\centerline{\epsfxsize=5.0in\epsfbox{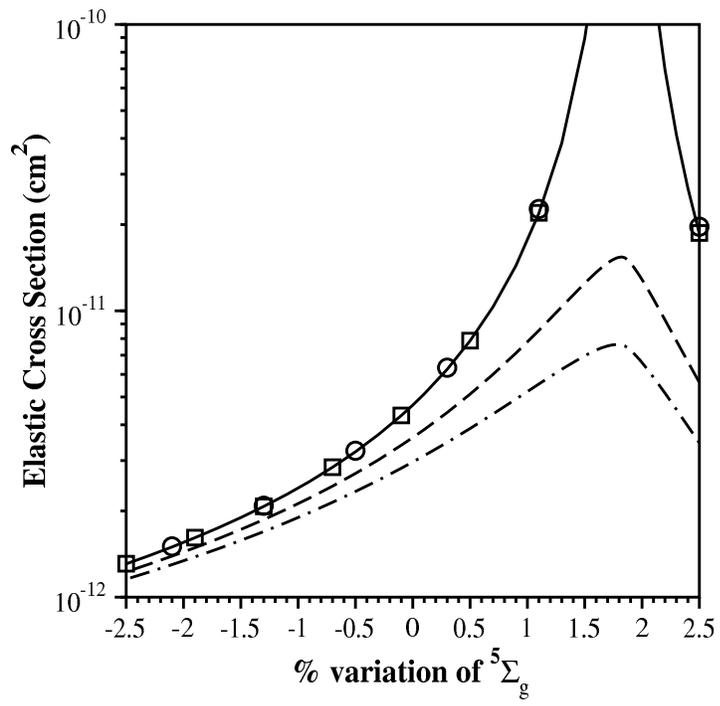}}
\caption{Thermally-averaged elastic cross sections for an unpolarized gas 
with curves labeled as per Fig.\ref{totalcce}. }
\label{totale}
\end{figure}

The elastic cross section depends on the real and imaginary scattering
lengths and its measurement in a spin-polarized or unpolarized gas may provide
useful information on $a_5$. In Figs. \ref{totalcce}--\ref{totalin} 
we provide the thermally-averaged 
total elastic cross sections and
Penning ionization rates for $(1,1)+(1,1)$ collisions and for an 
unpolarized gas calculated using Potential (A). 
Also shown are results for $1\,\mu $K and $500\,\mu $K 
calculated from the scattering
lengths using Eq.(\ref{rateeqnL}). In general the results obtained using 
Eq.(\ref{rateeqnL}) for temperatures up to $100\,\mu $K are identical to the
thermally-averaged results whereas at higher temperatures, 
outside the Wigner regime, the use
of scattering lengths is inappropriate and thermal averaging is required.
The rate equations
\begin{eqnarray}
\frac{\partial n_\alpha}{\partial t} &=& K^{\mathrm{inel}}_{\alpha,\alpha} \,
n^2_{\alpha} \nonumber \\
\frac{\partial n_\alpha}{\partial t} &=& K^{\mathrm{inel}}_{\alpha,\beta} \,
n_{\alpha} n_{\beta} 
\end{eqnarray} 
define our partial rates for,
respectively,
identical and non-identical atom
collisions, where $n_{\alpha}$ is the number of colliding atoms in state  
$\alpha $ and the superscript `$\text{inel}$' denotes 
`$\text{P}$', `$\text{ex}$' or `$\text{deg}$'.
The total thermally-averaged Penning rates and cross sections for an 
unpolarized gas 
are obtained assuming an equal population
of the $s=1$ magnetic substates so that
$n_{\alpha}=n/3$ and hence
\begin{eqnarray}
\frac{\partial n}{\partial t} &=& 
\frac{1}{9} \sum_{\alpha,\beta} \ 
\langle K^{\text{P}}_{\alpha,\beta}\rangle\;  n^2 
\nonumber  \\
\frac{\partial n}{\partial t} &=&\frac{1}{9} \sum_{\alpha,\beta} \ 
\frac{\langle v\sigma^{\mathrm{inel}}_{\alpha,\beta}\rangle}{\langle v \rangle}\;  n^2,
\end{eqnarray} 
where 
$\langle ... \rangle$ denotes the thermal average.
In this case the assumption that the magnetic substates are 
evenly populated in an
unpolarized gas  is well justified on collisional
grounds. 
At temperatures above $500\,\mu $K the degenerate rates $K^{\mathrm{imdeg}}$
evenly  mix $(1,1)$, $(1,0)$ and $(1,-1)$ atoms. At 
lower temperatures the Penning
rates $K^{\text{P}}_{(1,0),(1,0)}$ and $K^{\text{P}}_{(1,1),(1,-1)}$, 
which dominate the exothermic inelastic rates, 
deplete the three different atomic
populations $n_\alpha$ equally since 
$2 K^{\text{P}}_{(1,0),(1,0)}=K^{\text{P}}_{(1,1),(1,-1)}$ and the 
collision of
$(1,0)+(1,0)$ results in the loss of two $(1,0)$ atoms. Here we 
have neglected the small
contribution from spin-dipole processes, that is 
$K^{\text{ex}}_{(1,1),(1,-1)}$, and assume
that any initial asymmetry in the populations $n_\alpha$ due to 
preparation of the
atoms in a  light field for instance 
is small or has become small once the measurement of the 
collisional rate  in the absence of light is performed.
The thermally-averaged results were calculated by averaging over a 
Maxwell-Boltzmann
distribution of atomic velocity using 71 velocity nodes which correspond to
collision energies in the range $0.01\,\mu$K to $10,000\, \mu$K. Since the
results are for the case (A) potentials,
with the percentage variation in the
singlet potential set to zero, we estimate from the uncertainties in the
scattering lengths that the errors in the elastic cross sections and  
total inelastic rates are of
the order of 10\% and 40\% respectively. The Penning rates possess a larger
uncertainty to account for the percentage variation of the $^1\Sigma_g$ potential  
whereas the unpolarized elastic rates, which are dominated by the real 
scattering lengths belonging  to  collisions with $M=2$ or 1, 
 are controlled only by $^5\Sigma_g$.

In an unpolarized gas the $p$-waves can contribute  in  
$(1,1)+(1,-1)$,  $(1,1)+(1,0)$
and  $(1,-1)+(1,0)$ collisions. These  contributions  to the total 
thermally-averaged  Penning rates   were found to be  negligible at 1 $\mu$K. 
However the $p$-wave contributions increased  the total  Penning rate 
(compared to that obtained using only $s$-waves)  by  approximately  
7\% at 500 $\mu$K and 12\% at 1 mK. 
The $p$-waves modified the total elastic cross sections by less than 
1\% at all temperatures.

For $ (1,1)+(1,1)$ collisions (and similarly for $ (1,1)+(1,0),\; 
(1,-1)+(1,-1)$ or $(1,-1)+(1,0)$) we observe a resonance in the inelastic
rates at a percentage variation of +1.875 due to the resonant enhancement of the
exothermic rates. We find that $K^{\text{P}}_{\alpha,\beta} 
> K^{\text{ex}}_{\alpha,\beta}$, indicating that 
most but not all of the flux leaving the $[(1,1)+(1,1)],l=0$ entrance channel
is subsequently lost through ionization. These rates are much smaller
than those from the $(1,0)+(1,0)$ and $(1,1)+(1,-1)$ collisions and the  
total contribution to $K^{\text{P}}_{\alpha,\beta}$ from
$(1,1)+(1,1),\; (1,1)+(1,0),\; (1,-1)+(1,-1)$ and $(1,-1)+(1,0)$ 
collisions is only observable in Fig. \ref{totalin} as a
small peak at +1.875 in the unpolarized ionization rate.  

The total elastic cross sections of an unpolarized or a  polarized gas show 
strong dependences on the form of the ${}^5 \Sigma_g^{+}$ potential and
provide  possible measures of $a_5$.

\section{Conclusions}

The scattering lengths associated with the three molecular potentials relevant
to collisions of metastable helium atoms have been reported. The uncertainties
in the molecular potentials and autoionization widths have been considered and
probable ranges of values given for the scattering lengths for each molecular
state.
Scattering lengths for collisions involving the various atomic states have
also been calculated and related to the elastic cross sections and inelastic
collision rates for temperatures in the Wigner threshold region, with the aim
of providing a correspondence with experimentally measurable quantities. In
particular, it has been shown that measurement of the total elastic cross
section in a polarized or unpolarized gas should provide a means of
experimentally determining the $a_5$ scattering length, which is of importance
in the attainment of a Bose-Einstein condensate in a gas of spin-polarized
metastable helium atoms.

\begin{figure}[tb]
\centerline{\epsfxsize=5.0in\epsfbox{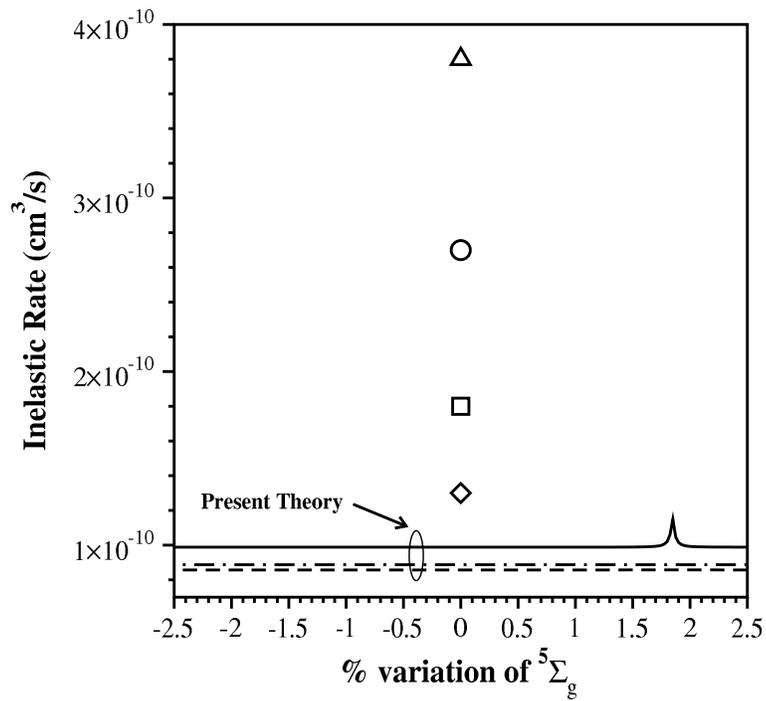}}
\caption{Thermally-averaged Penning rates for an unpolarized gas with 
curves labeled as per Fig.\ref{totalcce}. 
The theoretical predictions possess an
error of $\approx$ 40\% and the experimental results have 
uncertainties on the
order of 50\% Experimental results are denoted by 
$\bigtriangleup $ for \protect\cite{Kumakura&Morita:1999}, 
$\bigcirc $ for \protect\cite{Mastwijketal:1998a},
$\Box $ for \protect\cite{Toletal:1999} and
$\Diamond $ for \protect\cite{Herschbachetal:2000b}.}
\label{totalin}
\end{figure}

In Fig.~\ref{totalin} we compare the total Penning rates
for an unpolarized gas
calculated here with those from experiment. 
Not shown are the theoretical uncertainties of $\approx$ 40\% 
which arise from uncertainties in the
molecular potentials and in the form of the ionization widths. 
The total elastic
cross sections and Penning rates are consistent with those reported in
Ref.~ \cite{Venturi&Whittingham:2000} where  sightly different  molecular
potentials and ionization widths were used. The experimental results possess 
uncertainties on the
order of 50\% which are not shown in Fig.~\ref{totalin}. 
The experimental results correspond
to the case of zero magnetic field whereas the theoretical predictions 
are made for $B=10$~G.
However the scattering lengths were found to vary by less than 1\% 
over the range 0--20~G which
is negligible when compared to these uncertainties that arise 
from the form of the  ionization width. 
  The comparison between 
theoretical and experimental data  is satisfactory given these 
uncertainties, however the experimental
results are consistently higher than the theoretical predictions.

Prior to the submission of this paper, no experimental results or 
theoretical predictions existed for the 
scattering lengths,
cross sections and rates calculated here for incident atoms in specific
states. The $(1,1)+(1,1)$ spin-polarized system has been investigated by
Shlyapnikov {\it et~al.} \cite{Shlyapnikovetal:1994,Fedichevetal:1996a}
and \cite{Venturietal:1999} but no (quantitative) 
${}^{5}\Sigma_g^{+}$ scattering lengths were reported.
However, during the review of this paper, two measurements of scattering
length have been announced; $377 \pm 189\,a_{0}$ by \cite{Robertetal:2001}
and $302 \pm 151\,a_{0}$ by \cite{Pereiraetal:2001c}. These measurements, 
together with measured suppression 
by a factor of $> 2 \times 10^{3}$ 
for the Penning ionization rate for a spin polarized gas compared to that
of an unpolarized gas, are consistent with the current predictions.

Finally, using the scattering lengths reported in this investigation, one can
estimate the scattering lengths for the other isotopes of helium by mass
scaling the vibrational defect. This is related to the scattering length by
\cite{Mies}
\begin{equation}
a_{2S+1} = - \left. \frac{\partial \nu}{\partial \kappa}
\right|_{\kappa \rightarrow 0} 
\left[ \cot( \frac{\pi}{t-2}) + \cot(\nu_S(0)) \right] ,
\end{equation}
where $\nu_S(0)$ is the vibrational defect and $t$ is defined by the 
leading term $-C_t/R^t$ in  the long-range potential ($t=6$ for He). The 
term $\frac{\partial \nu}{\partial \kappa}|_{\kappa \rightarrow 0} $ 
is an asymptotic property which depends only on the long-range potential
and can be approximated by 
$0.956\, \times \, 0.5 (2 \mu C_6)^{0.25} \approx 35 $ for He \cite{Geback}. 
To mass scale the vibrational defect we first calculate
$\nu_S(0)$ for ${}^4$He for a given potential. Since the trigonometric
function is periodic, this only gives the fractional part of the 
vibrational defect and one must include the multiple of $n \pi$ where $n$ is
the number of bound states supported by that potential,
ie $\nu_S(0) \rightarrow n \pi +\nu_S(0) $. This vibrational defect 
can then be scaled using
$(\mu_{x}/ \mu_{4})^{0.5} \times  \nu_S(0)$,
to determine the vibrational defect for isotope $x$. Here $\mu_{x}$ and
$\mu_{4}$ are the reduced masses of ${}^{x}\text{He}$ and ${}^{4}\text{He}$
respectively.

\acknowledgments

VV acknowledges partial support from the Engineering and Physical
Sciences Research Council.  The Institute for Theoretical Atomic and
Molecular Physics is supported by a grant from the NSF to the Harvard
College Observatory and the Smithsonian Astrophysical Observatory.

\bibliographystyle{prsty}



\end{document}